\documentclass[conference]{IEEEtran}
\IEEEoverridecommandlockouts
\usepackage{natbib} 
\setcitestyle{numbers,square}
\usepackage{amsmath,amssymb,amsfonts}
\usepackage{algorithmic}
\usepackage{graphicx}
\usepackage{textcomp}
\usepackage{xcolor}
\usepackage{graphicx}
\usepackage{subfloat}
\usepackage{amssymb}
\usepackage{latexsym}
\usepackage{subcaption}
\usepackage{url}
\usepackage{amsmath}
\usepackage{color}
\def\BibTeX{{\rm B\kern-.05em{\sc i\kern-.025em b}\kern-.08em
    T\kern-.1667em\lower.7ex\hbox{E}\kern-.125emX}}
\begin{document}

\title{Application of attention-based Siamese composite neural network in medical image recognition\\

}

\author{\IEEEauthorblockN{1\textsuperscript{st} Zihao Huang*}
\IEEEauthorblockA{\textit{College of Computer Science and Electronic Engineering} \\
\textit{Hunan
University}\\
Changsha, China \\
zihaohuang@hnu.edu.cn}
*Corresponding author
~\\
\and
\IEEEauthorblockN{2\textsuperscript{nd} Yue Wang}
\IEEEauthorblockA{\textit{College of Computer Science and Electronic Engineering} \\
\textit{Hunan
University}\\
Changsha, China \\
nxlwwy@hnu.edu.cn}

~\\
\and
\IEEEauthorblockN{3\textsuperscript{rd} Weixing Xin}
\IEEEauthorblockA{\textit{College of Material Science and Engineering} \\
\textit{Hunan
University}\\
Changsha, China \\
lstar@hnu.edu.cn}
~\\
\and
\IEEEauthorblockN{4\textsuperscript{th} Xingtong Lin}
\IEEEauthorblockA{\textit{College of Computer Science and Electronic Engineering} \\
\textit{Hunan
University}\\
Changsha, China \\
inperson@hnu.edu.cn}
~\\
\and
\IEEEauthorblockN{5\textsuperscript{th} Huizhen Li}
\IEEEauthorblockA{\textit{College of Computer Science and Electronic Engineering} \\
\textit{Hunan
University}\\
Changsha, China \\
Joney@hnu.edu.cn}
~\\
\and
\IEEEauthorblockN{6\textsuperscript{th} Haowen Chen*}
\IEEEauthorblockA{\textit{College of Computer Science and Electronic Engineering} \\
\textit{Hunan
University}\\
Changsha, China \\
hwchen@hnu.edu.cn}
*Corresponding author

\and
\IEEEauthorblockN{7\textsuperscript{th} Yizhen Lao*}
\IEEEauthorblockA{\textit{College of Computer Science and Electronic Engineering} \\
\textit{Hunan
University}\\
Changsha, China \\
yizhenlao@hnu.edu.cn}
*Corresponding author

~\\
\and
\IEEEauthorblockN{8\textsuperscript{th} Xia Chen}
\IEEEauthorblockA{\textit{School of Basic Education} \\
\textit{Changsha Aeronautical Vocational and Technical College}\\
Changsha, China \\
xiachen4427@163.com}
}

\maketitle

\begin{abstract}
Medical image recognition often faces the problem of insufficient data in practical applications. Image recognition and processing under few-shot conditions will produce overfitting, low recognition accuracy, low reliability and insufficient robustness. It is often the case that the difference of characteristics is subtle, and the recognition is affected by perspectives, background, occlusion and other factors, which increases the difficulty of recognition. Furthermore, in fine-grained images, the few-shot problem leads to insufficient useful feature information in the images. Especially for rare diseases or in the early stages of an epidemic, when there is a lack of shared case databases among various medical institutions, the sample size can be extremely small, possibly less than 20. Considering the characteristics of few-shot and fine-grained image recognition, this study has established a recognition model based on attention and Siamese neural network. Aiming at the problem of extremely few-shot samples(samples $\leq$ 20), a Siamese neural network suitable for classification model is proposed. The Attention-Based neural network is used as the main network to improve the classification effect. Covid - 19 lung samples have been selected for testing the model. The results show that when the number of image samples is less than 20, the advantages of this composite model over an ordinary neural network are very pronounced. It means that this composite model is suitable for the recognition of extremely few-shot in fine-grained medical image analysis.
\end{abstract}


\section{Introduction}
In recent years, big data training for image recognition has been studied in depth. Many classical convolutional neural networks (such as Resnet18) have been widely used in the field of image recognition. However, these classical models require a large number of samples. It is often inadequate in practical applications since only insufficient set of samples are available. These include medical image analysis, crop disease identification, as well as applications in the fields of military and finance. Under the condition of small set of samples, deep learning suffers from over-fitting, low recognition accuracy and insufficient robustness. Therefore, few-shot learning is one of the challenges that image recognition often has to face. On the other hand, Computer vision and artificial neural network algorithms have been widely used in identifying large categories of objects, such as cat and dog classification, vehicle roadblock recognition, handwritten numerals recognition, face recognition, etc.\citep{tang2019random,luo2014switchable,fan2020sne,yoo2019pseudoedgenet}, in which the subtle differences between sub-classes make it easier to identify large classes of objects, while it is more difficult to identify more refined sub-classes. The difficulties come from the small visual differences between different subcategories under the same large category. When image recognition is affected by interference factors, it is more difficult to make correction identifications. Therefore, fine-grained image recognition is a challenge that intelligent image recognition often faces.

The applications of deep learning in image recognition, such as in medical image analysis, often face both challenges identified above. Typically, in the field of medical image analysis, computer-aided diagnosis systems can be used to improve the diagnostic accuracy of human experts\citep{zheng2018survey}, and may even replace human experts in the future. At present,  owing to the fact that many images are not in digital format, quality of images varies from different sources, and data annotation consumes a lot of time and labor, the standard sample data required for the learning model are insufficient, which is especially true in the cases of rare diseases, diseases not yet well understood, and epidemics at its early stage. In cases like these, the number of samples available is limited, resulting in enormous amount of difficulties in medical image recognition. Moreover, many medical image analyses depend on subtle differences in the images to diagnose subtle changes in the early stage of the disease or distinction between different diseases of the same organ, and varying degrees of noises of images further increase the difficulty of medical image recognition. In all, few-shot and fine-grained image recognition are two main challenges faced by deep learning technology in the application of medical image analysis.

For the problem of few-shot learning, Bromley, J \textit{et al.}\citep{bromley1993signature} proposed Siamese neural network. Siamese neural network contains two or more sub-networks with the same parameters and weights. Parameter updating can be carried out jointly on the two sub-networks. Sub-networks share weights so that training does not require a large number of parameters. Therefore, it has advantages in identifying small sample image sets and is less prone to over-fitting in recognition. This study is based on Siamese neural network, and its structure is modified to establish recognition model.

Traditional neural network features equal extraction of image characters, both in the image channel and in spatial plane. But when people make observations, they typically have elements they lend focus on. The direction of the development of deep learning is to simulate the 'selectability and differentiability' of human observers. This is the 'Attention Model' proposed in image recognition research. This model can selectively extract features from an image and solve the problem of small differences between image sub-classes, and is greatly affected by perspectives, backgrounds, occlusion, noise and other factors\citep{mnih2014recurrent}. Therefore, the image recognition model established in this study uses the “attention mechanism” principle.

To summarize, this study has established a new composite deep learning model based on attention mechanism and Siamese neural network model. First of all, we establish a classifiable siamese neural network. Then, the effects of different attention mechanisms on different networks are tested, and finally the model with the best performance is selected to be embed into the classifiable Siamese neural network to improve the model effect. Covid-19 lung samples are used for testing the model.

The experimental results show that the composite deep learning model developed in this study has the following advantages. 
\begin{itemize}
\item It solves the problem that a large number of samples are needed in deep learning in the traditional models. The established model shows obvious advantages in identifying small sample image sets such as Covid-19. The fewer the samples are, the better the recognition effect is compared with the traditional network. 
\item The problem of feature extraction without center of gravity in deep learning is solved. The lungs infected by Covid-19 can be identified from the lung images. With the attention mechanism added, the classification accuracy of the samples is improved.
\item The proposed lightweight composite model simplifies the calculation and reduces the requirements to the computing power. Covid-19 can be diagnosed rapidly and at a lower cost, which is significant in rapid diagnosis of epidemic diseases. It can be easily deployed.
\end{itemize}

\section{Related Work}

\subsection{Recognition of medical images}
Ever since scientists discovered X-rays, X-ray imaging has been an indispensable tool.  Doctors have been increasingly relying on medical imaging for diagnosis. However, doctors’ assessments on medical images have been, avoidably, to certain degree subjective. Different doctors can give different interpretations to the same image. Errors in this high-intensity work may also lead to false diagnosis. These motivates the development of computation-based technology to assist medical image recognition.

Since the outbreak of Covid-19 in 2019, related researches based on deep learning have made good progress\citep{wang2020novel}. The diagnosis by RT-PCR of suspected infection has received positive result. However, relevant studies have shown that the detection rate of RT-PCR is far from adequate, generally about 30 \% - 60 \%\citep{wang2020weakly}, and chest CT images can only be assisted to detect the disease to a certain extent. To improve the efficacy, a deep learning model might be useful to detect and classify the lung images of Covid-19 infection, assisting doctors to identify  'highly suspected pneumonia patients'. In 2020, Bai \textit{et al.} first segmented the CT image, and used 2DEfficientNetB4 to score the segmented image. With the idea of ensemble learning, the scores of many CT slices were integrated and the final prediction was made\citep{bai2020ai}. In the same year, Wang \textit{et al.} first used the U-Net segmentation model to obtain images of the lung region, and used the segmented images and the original images as the input of 3D deep neural network, and predicted the probability of COVID-19 infection.

\subsection{Few-Shot Learning and Siamese Neural Network}
In recent years, deep learning technology has seen abundant research results in the field of image recognition. However, most of the current technologies require the support of a large number of samples, which limits the application of deep learning technology in cases where only small number of samples are available. These include some cases in medical fields, astronomy, and identification of endangered animals and plants, to name a few. At present, the solutions for few-shot learning include transfer learning, data augmentation and meta-learning\citep{long2018transferable,perez2017effectiveness,vilalta2002perspective}.

However, transfer learning has high requirements for the source domain samples. If the selection is incorrect, it may produce negative effect. Besides, for the second method, the effectiveness of data augmentation for improving the fine-grained image recognition is limited. It may increase the impact of noise. The main purpose of meta-learning is to learn a universal and generalized representation method in data, which can be directly used in the target data set,  such as in the learning “comparative ability” in Siamese neural networks. Siamese neural network was first proposed in 1993. In 2005, Chopra \textit{et al.} proposed a network to train “comparative ability” from data\citep{chopra2005learning}.

\subsection{Fine-Grained Image Recognition and Attention Mechanism}
At present, there are two mainstream methods for fine-grained image recognition. One is the recognition model based on strong supervision information and the other is the recognition model based on weak supervision information.\citep{zhang2014part,shih2015part,branson2014bird,huang2016part}

In addition to the two methods mentioned above, an attention mechanism has also been proposed. Through task-oriented screening and filtration of information, human attention mechanism enables our brain to focus on processing the information related to tasks, thereby improving the efficiency of the information processing and utilization. The attention mechanism in computation was first proposed in the field of computer vision. Although the essential concept has long been proposed, it gained real attention only when the Google DeepMind team used the concept in image classification in June 2014. In the same year, Bahdanau \textit{et al.} used the concept in machine translation\citep{bahdanau2014neural}. Following these developments, attention mechanism has been widely adopted in deep learning. In the field of computer vision, there are several classic attention mechanisms. SEnet proposed in 2017 laid the foundation for the development of subsequent channel attention mechanisms\citep{hu2018squeeze}. SKnet and ECAnet are developed on this foundation\citep{li2019selective,wang2020supplementary}. In 2019, Li \textit{et al.} proposed a lightweight attention mechanism SGEnet, which also combines spatial and channel attention mechanisms\citep{li2019spatial}.

\section{METHDOLOGY}
Firstly, we have studied the effects of different attention mechanisms on classical neural networks. Secondly, using the idea of Siamese neural network, we have constructed a classified Siamese neural network model, embedded the neural network with attention mechanism which preforms well, and tested the effectiveness of the model. After that, we have gradually reduced the sample size, and compared to the effects of without  using Attention-based Siamese neural network in order to explore the role of Attention-based Siamese neural network in few-shot learning. In the end, we build an Attention-Based Siamese Neural Network and achieved good recognition accuracy in the identification of Covid-19 lung.

\subsection{Attention-Based Neural Network}
First of all, we need to make assessment on the best combination of attention mechanism and different neural networks.

Two classical neural networks, InceptionV3 and Resnet18, are used as the foundations, in which some attention mechanisms are respectively embedded: SE, SK, SGE and ECA. Since an attention mechanism does not change the shape of the feature map, it can be placed in any convolution layer of a neural network. In order to save computation resources, and also avoid destroying the initial network structure, we have designed a composite attention-based neural network.

The inceptionV3 consists of 11 blocks of five types. Each block consists of many convolution layers and pooling layers, and the model is relatively complex.  Therefore, it should be avoided to add too many attention modules. Even with one attention module for each block, we would have 11 new modules, a number that still could be too large. Therefore, we decided to add one attention module to each type of blocks. Finally, we add five attention modules. 

For Resnet18, in addition to the ordinary convolution layer and pooling layer, the most important thing is to include two residual blocks. Because a residual block is a structure of skip connections, in order to maintain the integrity of such a structure, the attention module should not be placed inside a residual block. It can be seen that each two residual blocks can be regarded as a whole. Therefore, we consider each two residual blocks as a layer, a total of four layers, and add the attention module at the end of each layer. Four attention modules are added.

The function that the attention-based neural network needs to complete is to perform multiple classifications of images. We have used a cross entropy loss function to describe it. Because the samples are four categories, the four-classification cross entropy loss function is described by the following:

\begin{equation}
    L=\frac{1}{N} \sum_{i} L_{i}=-\frac{1}{N} \sum_{i} \sum_{\mathrm{c}=1}^{4} y_{i c} \log \left(p_{i c}\right)
\end{equation}
Where $N$ is the number of samples, and $y_ic$ is either 0 or 1. If the real category of sample $i$ is equal to $c$, then $y_ic$ is 1, otherwise 0. $p_ic$ is the probability that the observation sample $i$ belongs to category $c$.

\subsection{Classifiable Attention- Based Siamese Composite Neural Network}
\subsubsection{Establishment of Attention-Based Siamese Composite Neural Network}
Furthermore, we have integrated the attention-based neural network into a classifiable Siamese neural network.

A classifiable Siamese neural network can be divided into two parts: a front-end contrast-training structure and a back-end classifying-predicting structure. The contrast-training structure accepts inputs of two images and outputs the spatial distance of the two images. The training process is to make the distance of the same category of images in space as small as possible, and the distance of different categories of images in space as large as possible. The back end receives the distance from the front end and outputs the category of an image. Its overall network structure has two main types, training structure and predicting structure (Figures 1-a and 1-b).
\begin{figure}[htbp]
\centering
\begin{subfigure}[b]{0.4\textwidth}
  \includegraphics[width=\textwidth]{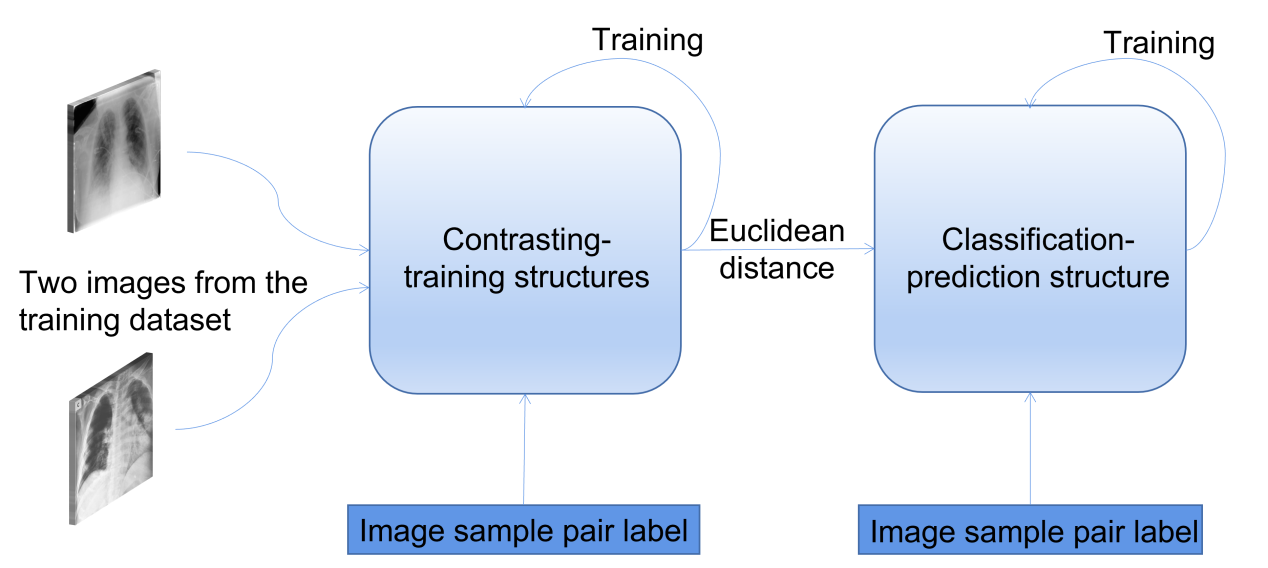}
  \caption{Classifiable Siamese Neural Network Training Structure}
\end{subfigure}
\hfill
\begin{subfigure}[b]{0.4\textwidth}
  \includegraphics[width=\textwidth]{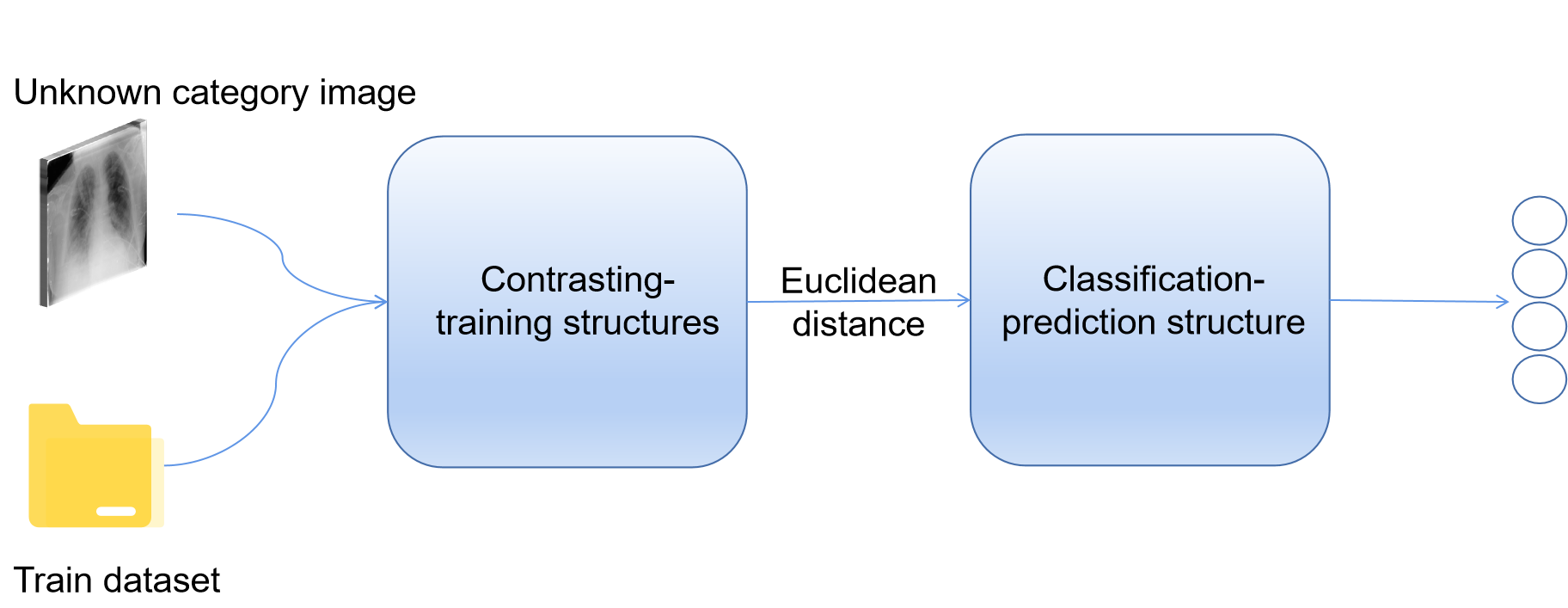}
  \caption{Classification Siamese neural network predicting structure}
\end{subfigure}
\caption{Classifiable Attention- Based Siamese neural network}
\end{figure}

As shown in Fig. 1-a, the training structure accepts a sample pair, all from the training set, and the front end and the back end are trained at the same time. In Fig. 1-b, the picture in the predicting structure is a picture to be classified, and the other picture is from the training set. Finally, the category is the output.

Front-end: contrast-training structure. This part is mainly composed of two attention-based neural networks with shared weights. As shown in Fig. 2, the two images pass through the neural network in the first part to obtain a vector, of which its value is the category number (here is 4). At the same time, the vector of 512 is obtained through a fully connected structure. The distance of the two images can be obtained by taking the Euclidean distance of the two vectors. (see Fig. 2-a and Fig. 2-b).

\begin{figure}[!t]
\centering
\begin{subfigure}[b]{0.4\textwidth}
  \includegraphics[width=\textwidth]{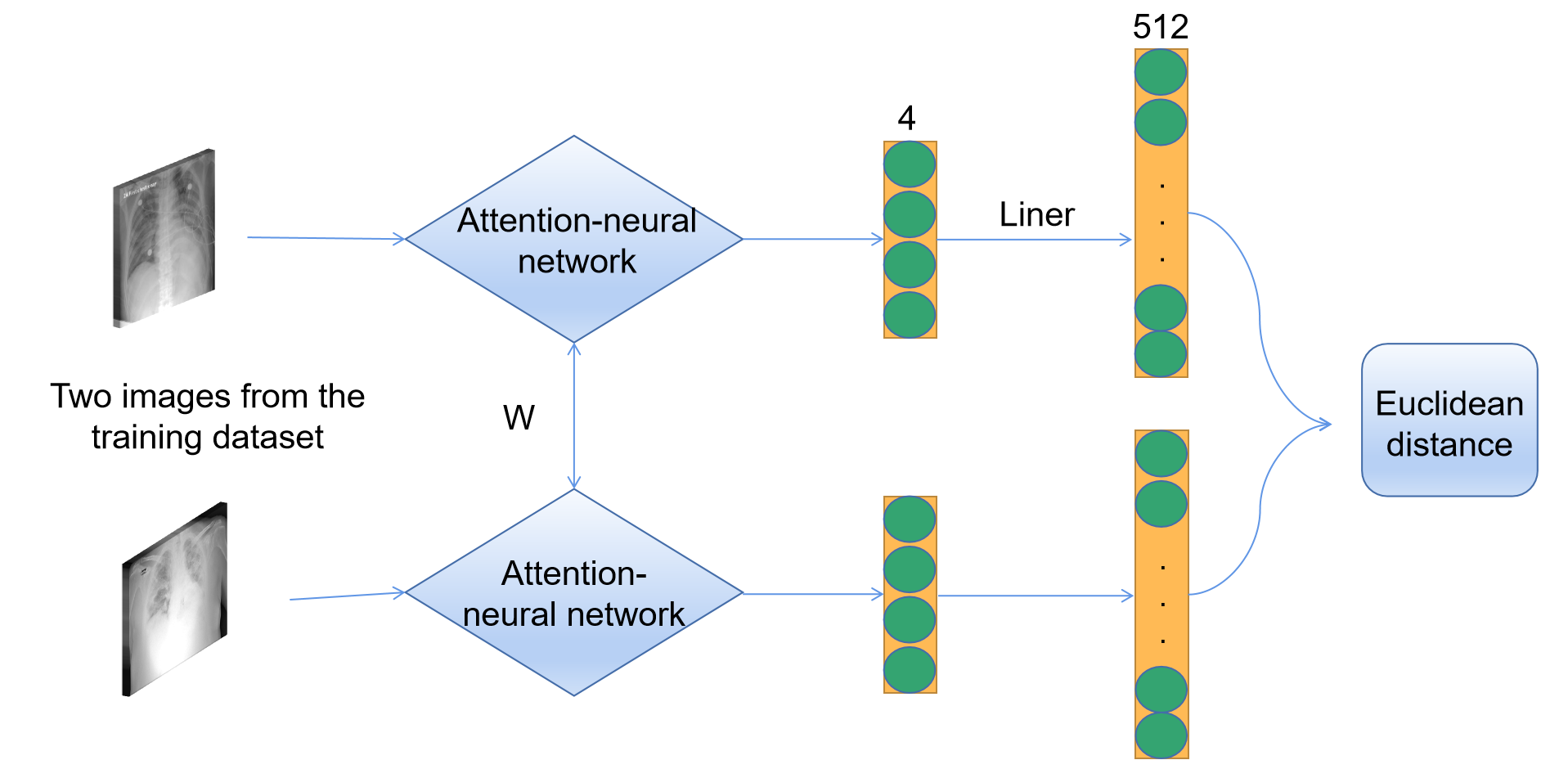}
  \caption{Contrast-training structure in training structure}
\end{subfigure}
\hfill
\begin{subfigure}[b]{0.4\textwidth}
  \includegraphics[width=\textwidth]{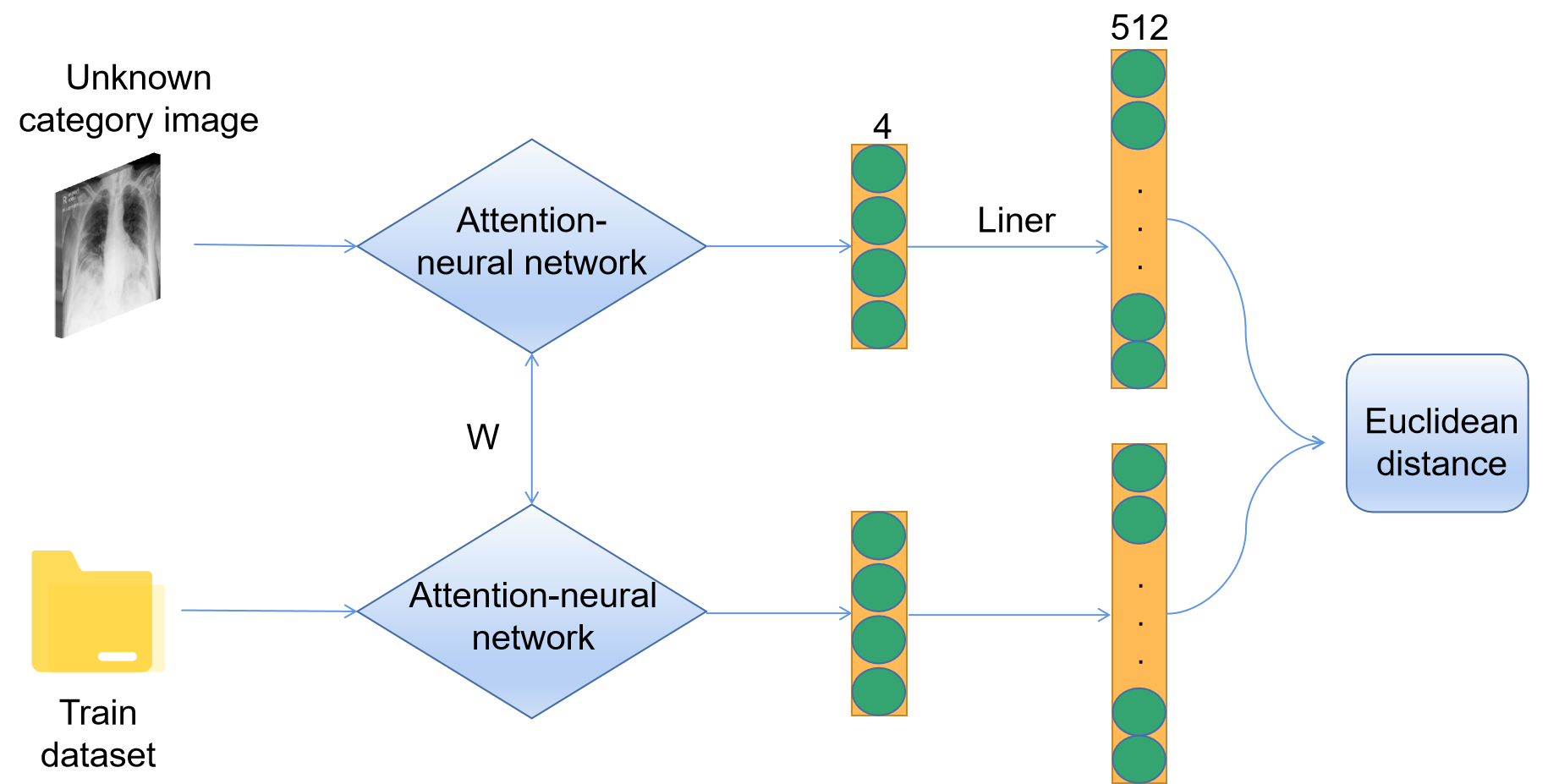}
  \caption{Contrast-training structure in predicting structure}
\end{subfigure}
\caption{Contrasting-training structure}
\end{figure}

Fig. 2-a shows the contrast-training structure in the training structure. The model receives the constructed sample pairs and obtains the spatial distance through the forward propagation of the model. This distance is combined with the loss function, and the reverse propagation is used for training. Fig. 2-b shows the contrast-training structure in the predicting structure. An unknown picture input by the model is fixed. Another picture comes from the training set, the output Euclidean distance is passed into the classifying-predicting structure to help to complete the classification. If the classification cannot be completed, an image is selected from other categories of the training set to construct the sample pair until the back end completes the classification.

Backend: classifying-predicting structure. When we use the training set to train, each iteration will output the distance of all the sample pairs, and each sample pair will have a label 0 or 1. The purpose of the back end is to map the distance to the probability of belonging to the same category. When one of our sample pairs belongs to the same category, the label is 0. Therefore, the problem is equivalent to the probability of predicting the label of a sample pair composed of an unknown sample and a known sample to be zero. ( See Fig. 3-a and Fig. 3-b ).

\begin{figure}[htbp]
\centering
\begin{subfigure}[b]{0.3\textwidth}
  \includegraphics[width=\textwidth]{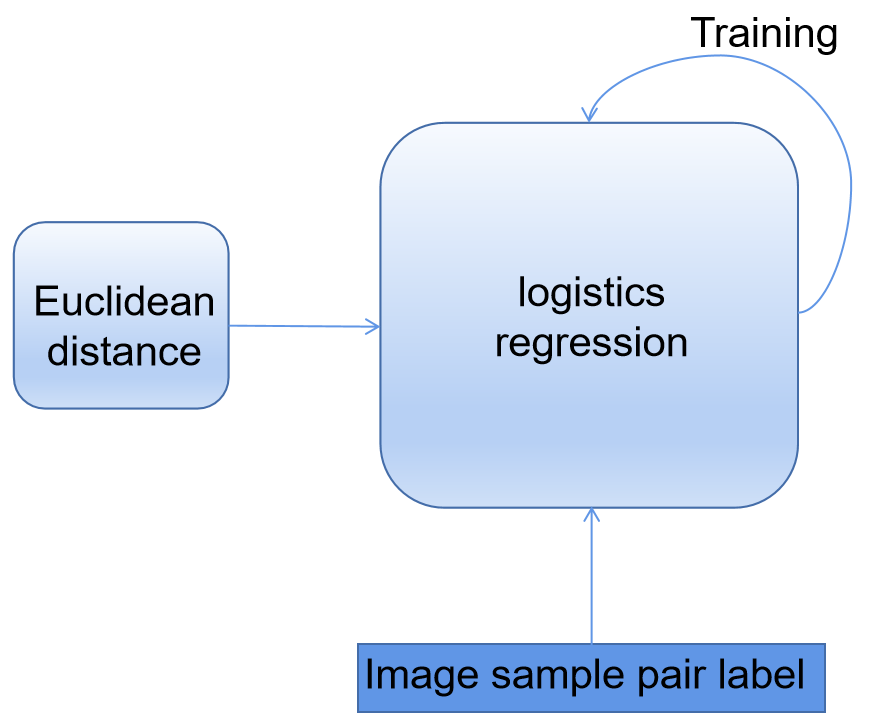}
  \caption{Classifying-predicting Structure in Training Structure}
\end{subfigure}
\hfill
\begin{subfigure}[b]{0.4\textwidth}
  \includegraphics[width=\textwidth]{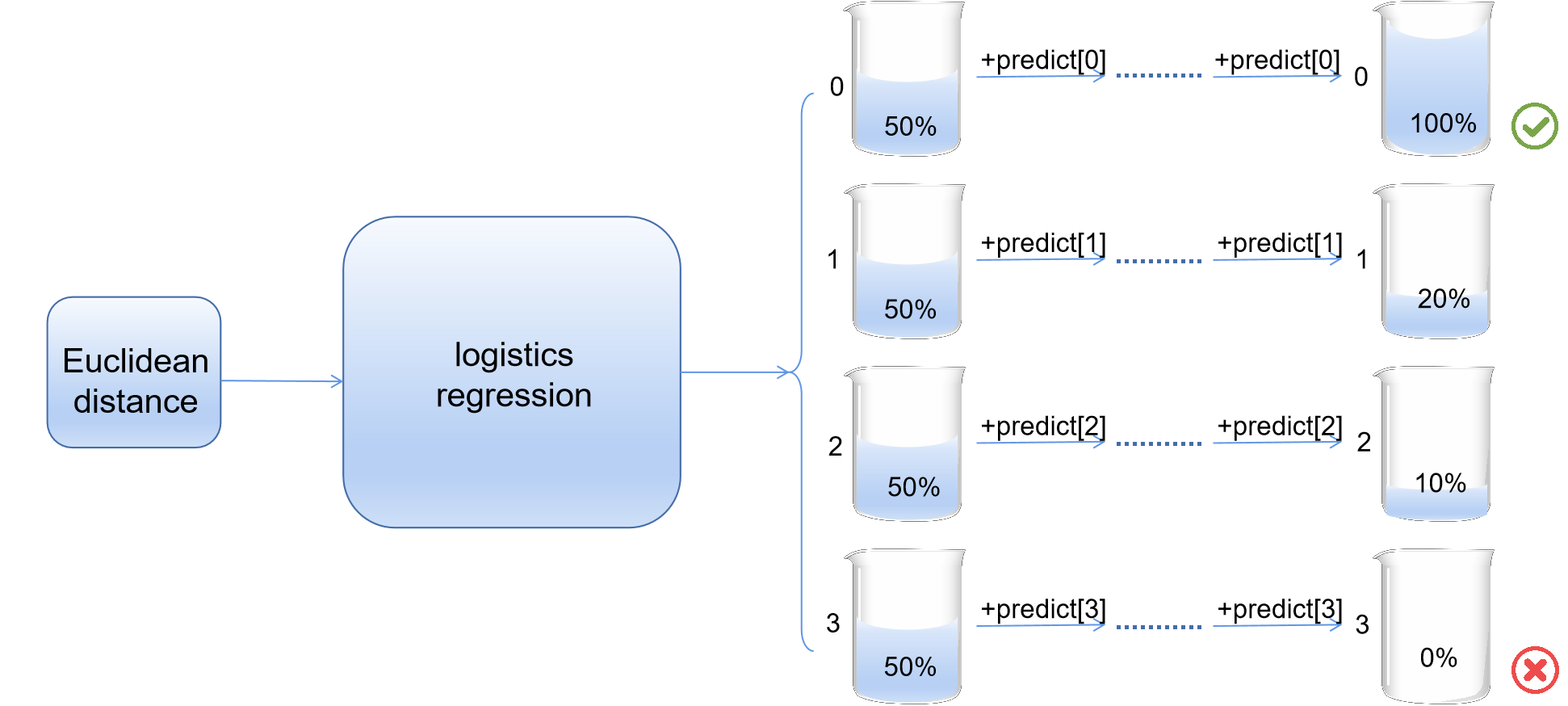}
  \caption{Classifying-predicting structure in predicting structure}
\end{subfigure}
\caption{Classifying-predicting Structure}
\end{figure}

As shown in Figure 3-a, in order to ensure the accuracy of a prediction and the real-time nature of the prediction, we use a logistic regression model. The distance obtained from each front-end training is taken as the back-end sample feature, and the back-end sample label is same as the front-end sample label, which is put into the logic regression model for learning. For prediction, it is unreasonable to determine whether an unknown picture belongs to this category by only one known picture. Therefore, we design a model shown in Fig. 3-b, where we set four basic probabilities base[i]. In the experiment, the initialization is set to 0.5, representing the probability of four categories. We continuously use the Euclidean distance output by the front end as the input, and iteratively update the probability in the four categories. The formula of probability iteration is as the following:
\begin{equation}
    \operatorname{base}[i]+=\operatorname{base}[i]+(\operatorname{predict}[i]-0.5)
\end{equation}
Among them, prediction[i] denotes the probability between the known class i image and an unknown image passing through the back-end logic regression output. When the probability of a certain category exceeds 1, the sample is determined to belong to this category. If the probability of a certain category is less than 0, the sample does not belong to this category, and the probability iteration of this category is abandoned.

\subsubsection{ Loss Function of Classifiable Attention-Based Siamese Neural Network}
Our objective is to train the distance of images belonging to the same category as small as possible. On the contrary, the distance of images belonging to different categories in space is as large as possible. Therefore, the loss function can choose the contrastive loss function\citep{hadsell2006dimensionality}. The contrastive loss function formula is:
\begin{equation}
    L=\frac{1}{2 N} \sum_{i=1}^{N}\left((1-y_{i})d_{i}^{2}+ y_{i} \max (\operatorname{margin}-d_{i}, 0)^{2}\right)
\end{equation}

where $d_{i}$ represents the distance between two vectors, which is the output of the front end; $y_{i}$ represents the label of a sample pair; 0 represents that two images belong to the same category; 1 represents that two images do not belong to the same category. Margin represents the threshold of the distance. When a sample belongs to different categories, the distance exceeds the threshold, and the loss is set to 0. In our algorithm, the margin is set to 2. From the above contrastive loss function, when samples belongs to the same category, the loss function is $\sum_{i=1}^{N} y_{i} d_{i}^{2}$ . Therefore, with the learning of neural network, the loss value decreases continuously, and the distance between the samples decreases continuously as well. When samples does not belong to the same category, the loss function is $\sum_{i=1}^{N}(1-y_{i}) \max (\operatorname{margin}-d_{i}, 0)^{2}$,the loss value decreases continuously, the distance $d_{i}$ is also increasing. These are consistent with our expectation of using distance to evaluate whether samples belong to the same class.

\subsection{Model Evaluation Indicators}
In the experiment, the model is a four-classification problem, and the kappa coefficient is used for evaluation . The calculation formula is as the following:
\begin{equation}
    k=\frac{p_{o}-p_{e}}{1-p_{e}}
\end{equation}

Among them,  $p_{o}$is the sum of the correctly classified samples of each class divided by the total number of samples, which is simply our accuracy. The calculation of $p_{e}$ is given by:
\begin{equation}
    p_{e}=\frac{a_{1} \times b_{1}+a_{2} \times b_{2}+\ldots+a_{C} \times b_{C}}{n \times n}
\end{equation}
Where ,$a_{1}, a_{2} \ldots a_{C}$ is the number of real samples for each class, and $b_{1}, b_{2} \ldots b_{C}$ is the number of samples predicted for each class.  is the total number of samples. 
Kappa calculation results are between -1 ~ 1, the greater the Kappa coefficient is, the better the model classification effect is.

\subsection{Experiment}
In this section, we will describe the hardware we use, the data set, and the details and results of the experiment.
All experiments are completed by using pytorch 1.9.0. In terms of hardware, two Titan XP from nvidia are used for GPU acceleration, and the CUDA version is 11.2.
\subsection{Data Sets}
In this experiment, we use one dataset, COVID-19 Radiography Database\citep{chowdhury2020can,rahman2021exploring}.

COVID-19 Radiography Database: A team of researchers from Qatar University, Doha, Qatar, and the University of Dhaka, Bangladesh along with their collaborators from Pakistan and Malaysia in collaboration with medical doctors have created a database of chest X-ray images for COVID-19 positive cases along with Normal and Viral Pneumonia images. The dataset is still updated, initially containing only 219 COVID-19,1341 normal lung images and 1345 viral pneumonia chest images. Up to now, the dataset has 3616 COVID-19 lung images, 6012 Lung Opacity, and 1345 Viral Pneumonia images and corresponding lung masks. The dataset also reflects the problem of sample scarcity in the early outbreak of an infectious disease. It also proves the necessity of few-shot learning. 

\subsection{Experiment Detail}
\subsubsection{Attention-Based Neural Network}
In order to test the best combination of different attention mechanisms and different networks, we have done the following experiments.

\textbf{COVID-19 Radiography Database}: Because we only needed to test the effectiveness of the attention mechanism, we streamlined the dataset, and in order to balance the data, we decide to keep 400 images randomly for each category.

\textbf{Division of Data Sets}: To evaluate the ability of model learning and prediction, the dataset was divided into a training set of 300 images and a testing set of 100 images. All training sets account for 75 \% of the total number of samples. The number of samples are more balanced in both the training set and the testing set. This improves the reliability of the model assessment.

\textbf{Experiment Process and Results}

Next, we complete the fusion of two networks and four attention mechanisms, and test them on the constructed testing set. The final results take the optimal kappa value of the testing set (Table 1).

\begin{table}[htbp]
\caption{The results of attention-based neural network}
\begin{tabular}{ccc}
\hline
Attention mechanism & ResNet18 & InceptionV3          \\ \hline
None                & 0.688    & 0.792                \\
SEnet               & 0.816    & 0.868                \\
SK                  & \textbf{0.847}    & 0.875                \\
SGE                 & 0.795    & \textbf{0.882}                \\
ECA                 & 0.833    & 0.865                \\ \hline
\end{tabular}
\end{table}

As shown in Table 1, after adding the attention mechanism, the classification effect of InceptionV3 is slightly improved. Resnet18 is also improved.

\subsubsection{ Attention-based Siamese Composite Neural Network}

\textbf{Data Reduction}
Since the problem solved in our research is the few-shot problems, the training set should be reduced. We use two training sets, namely, the training set with only 10 pictures in each category and the training set with only 20 pictures in each category, and the test set still uses the test set used in the previous part.

\textbf{Construction of Dataset}
The dataset of Attention-based Siamese neural network is mainly divided into three categories: the training set during training, the test set during testing, and the training set that needs to be referred to during testing.

Training set during training: since each image can form a paired sample with any other image, the training set of 10 pictures in each category can form up to 39 + 38 + … + 1 = 780 paired samples, and the training set of 20 pictures in each category can form up to 79 + 78 + ... + 1 = 3160 paired samples. Therefore, the sample size (length) of the training set during training should be optimized, and the number of positive and negative samples should be uniform (Fig. 4).

\begin{figure}[htbp]
\centering
\includegraphics[scale=0.5]{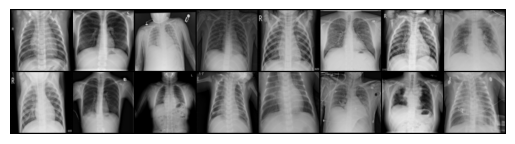}
\caption{Training samples}
\end{figure}

As shown in Fig.4, a training set of size 8 is constructed. Sample labels are [0,1,0,0,0,1,1,1], which contains 4 positive samples and 4 negative samples.
When testing, the testing set and the reference training set are made of single image. The batch size is fixed to 1. When testing, images one by one from the testing set and randomly paired with each type of image in the reference training set until the completion of the test.

\textbf{Experiment Process and Results}
First, we set the hyperparameters: In this section, the hyperparameters include the length of the training set, the batch size of the training set, the base value in the classification-prediction structure, the learning rate and the epochs. Because the batch size of the two data sets in the test part is set to 1, it is not a hyperparameter (Table 2).

\begin{table}[htbp]
\caption{The hyperparameter of different training set}
\begin{tabular}{lccccc}
\hline
                                   & length & batchsize & Base & lr & epochs \\ \hline
10 images & \textbf{500}    & 32        & 0.5  & 0.001         & 30     \\
20 images & \textbf{2000}    & 32        & 0.5  & 0.001         & 30     \\ \hline
\end{tabular}
\end{table}
As shown in Table2, all hyperparameters are consistent except for the length of the training set.

In order to illustrate the advantages of Attention-based Siamese network in solving the few-shot problems, we run the neural network only using attention mechanism and the attention-based Siamese neural network. we select the best two networks as the attention-based neural network in the previous section: Resnet18-SK, InceptionV3-SGE.

Firstly, the performance of each training set with 20 images per class in two modes and two kinds of neural networks is tested. we still use kappa coefficient as the evaluation index, which is the best under the testing set (Table 3).
\begin{table}[htbp]
\caption{Test results for a training set of 20 images per class on different networks}
\begin{tabular}{ccc}
\hline
Attention-Based neural network & \multicolumn{2}{c}{Kappa of Testing set} \\ \hline
\multicolumn{1}{l}{}           & No-Siamese           & Siamese           \\
Resnet18-SK                    & 0.334                & \textbf{0.563}             \\
InceptionV3-SGE                & 0.378                & \textbf{0.503}             \\ \hline
\end{tabular}
\end{table}

As shown in Table 3, with few-shot samples, all kappa values decrease, but the attention-based Siamese composite neural network is much more superior to the neural network with only attention mechanism in the previous part.

Next, we continue to reduce the number of training set images, and test the performance of training set with 10 images per class in six models (Table 4).

\begin{table}[htbp]
\caption{Test results for a training set of 10 images per class on different networks}
\begin{tabular}{ccc}
\hline
Attention-Based neural network & \multicolumn{2}{c}{Kappa of Testing set} \\ \hline
\multicolumn{1}{l}{}           & No-Siamese           & Siamese           \\
Resnet18-SK                    & 0.192                & \textbf{0.523}             \\
InceptionV3-SGE                & 0.298                &\textbf{0.363}             \\ \hline
\end{tabular}
\end{table}

As shown in Table 4, with the further reduction of each type of images, the neural network with only attention mechanism in the previous part has basically lost its learning ability, the model does not have the ability to predict, and the kappa value approaches 0. However, the classification ability of attention-based Siamese composite neural network has almost no change.

In addition to the best kappa coefficient of the testing set, we are also concerned about the training process of the model. We find that in few-shot problems, the stability of the training set and the speed of the model convergence are different in both modes (Fig. 5).

\begin{figure}[htbp]
\centering
\begin{subfigure}[b]{0.5\textwidth}
  \includegraphics[width=\textwidth]{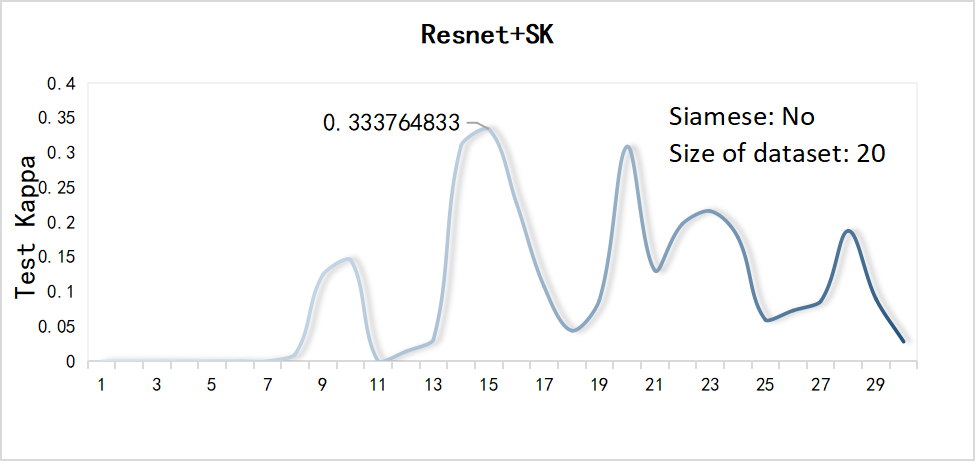}
  \caption{training process of the model without Siamese}
\end{subfigure}
\hfill
\begin{subfigure}[b]{0.5\textwidth}
  \includegraphics[width=\textwidth]{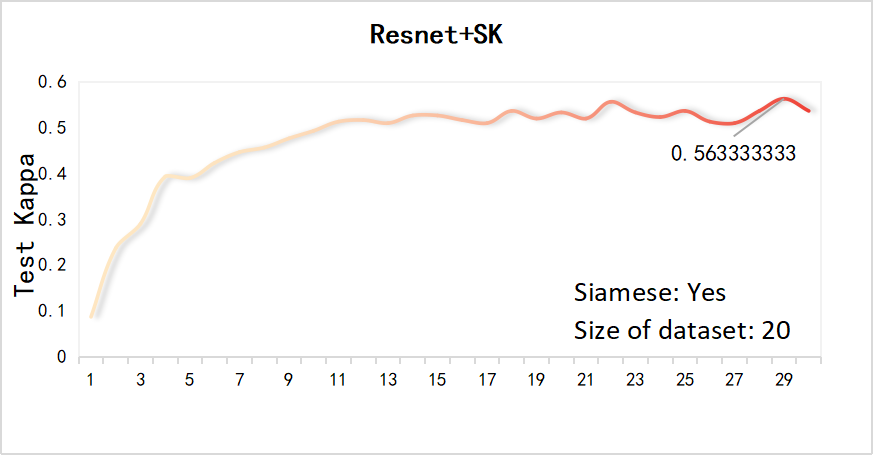}
  \caption{training process of the model with Siamese}
\end{subfigure}
\caption{training process of the model}
\end{figure}

Figure 5 shows the variation of Kappa with the number of iterations in Resnet + SK in 20 Covid-19 lung samples. It can be seen that when the Siamese neural network is not used, the convergence rate is slow, and the training fluctuation is remarkable, and the training Kappa peak is only 0.3338. However, after using the Siamese network, the convergence rate of the model is fast, the model fluctuation is slight, the training is more stable, and the training peak even is 0.5633.

\section{CONCLUSION}

This paper presents a composite model of attention-based Siamese neural network, which can be used to solve the problem of few-shot and fine-grained. Good performance is demonstrated in the application of Covid-19 lung identification. With a training set of only 10 images in each category, the traditional convolutional neural network is incapable of learning any useful features, cannot classify the unknown categories, and the kappa coefficient is essentially 0. On the other hand, with the same data set, the model proposed in this paper improves the kappa coefficient remarkably.

For the traditional convolutional neural network, with the increase of number of categories, the effectiveness of the model decreases. On the other hand, due to the attention-based Siamese neural network’s learning ability of comparison, the model has stronger learning ability under more categories of samples. We can expect more in-depth research on the relationship between category number and model learning ability in the future. Moreover, not only can this model be applied to the identification of Covid-19 lung, but also be extended to other few-shot and fine-grained image recognition fields.

\section*{Acknowledgment}

This research was funded by the Innovation and Entrepreneurship Training Program of Hunan Province under Grant S202110532360, the Natural Science Foundation of Changsha City under Grant KQ2202137, and the Changsha City takes the lead in major science and technology projects under Grant No. KQ2102002. The authors thank the anonymous reviewers and editors for their constructive comments on this manuscript.

\section*{References}
\bibliographystyle{IEEEtran}
\bibliography{IEEEabrv,refs.bib}
\end{document}